\pdfoutput=1

\documentclass[a4paper]{article}
\usepackage[a4paper]{geometry}
\usepackage{lineno,hyperref}
\usepackage{mathtools}
\usepackage{graphicx}
\usepackage{float}
\usepackage{amsthm}
\usepackage{amssymb}
\usepackage{amsmath} 
\usepackage{multirow}
\usepackage{multicol}
\usepackage{caption}
\usepackage{subcaption}

\usepackage[T1]{fontenc}
\usepackage{lmodern}

\usepackage{booktabs} 
\usepackage[linesnumbered,ruled]{algorithm2e}

\date{}
\title{Opportunistic multi-party shuffling for data reporting privacy}

\author{
	Marios Fanourakis\\
	\texttt{CUI, Quality of Life Lab}\\
	\texttt{University of Geneva, Switzerland}\\
	\texttt{marios.fanourakis@unige.ch}
}

\begin{document}
\maketitle

\begin{abstract}
		An important feature of data collection frameworks, in which voluntary participants are involved, is that of privacy. Besides data encryption, which protects the data from third parties in case the communication channel is compromised, there are schemes to obfuscate the data and thus provide some anonymity in the data itself, as well as schemes that 'mix' the data to prevent tracing the data back to the source by using network identifiers. This mixing is usually implemented by utilizing special mix networks in the data collection framework. In this paper we focus on schemes for mixing the data where the participants do not need to trust the mix network or the data collector with hiding the source of the data so that we can evaluate the efficacy of peer to peer mixing strategies in the real world. To achieve this, we present a simple opportunistic multi-party shuffling scheme to mix the data and effectively obfuscate the source of the data. We successfully simulate $3$ cases with artificial parameters and then use the real-world Mobile Data Challenge (MDC) data to simulate an additional $2$ scenarios with realistic parameters. Our results show that such approaches can be effective depending on the time constraints of the data collection and we conclude with design implications for the implementation of the proposed data collection scheme in real life deployments.
\end{abstract}


\section{Introduction} 
Mobile crowd sensing leverages the number of user-companioned devices, including mobile phones, wearable devices, and smart vehicles, and their inherent mobility to collect information such as location, personal and surrounding context, noise level, and more \cite{Guo2014}. The users, acting as sensors, have a certain expectation of privacy about the data they might be sharing and often do not trust that is it possible to hide their identity while at the same time provide usable data \cite{Gustarini2016a}. Providing data privacy in crowd sensing or other participatory data collection context has been an important task that ensures that the participants privacy is protected (ex., data cannot be traced to the individual) while the data is being collected at large scales without bias stemming from privacy-aspects (ex. participants switching off their phone in certain contexts). There are several elements of the data collection process that can be exploited to reveal sensitive information about the participants: the data communication channel, the reporting of the data, and the data itself.
\begin{figure}[H]
	\centering
	\includegraphics[width=0.5\textwidth]{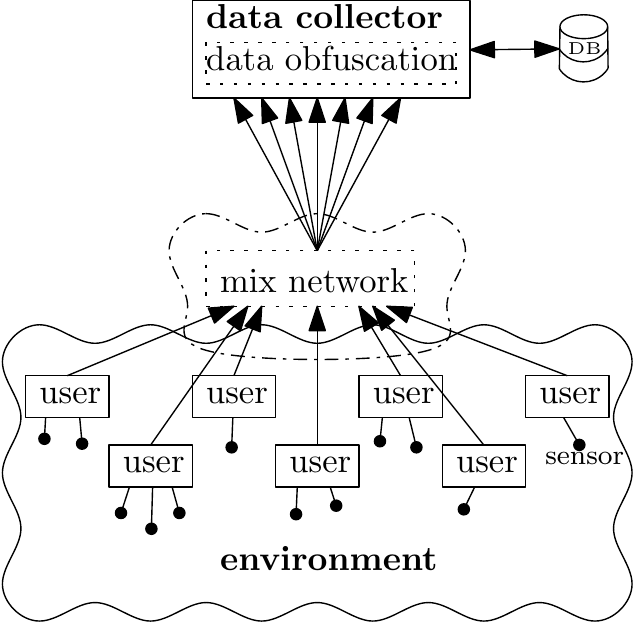}
	\caption{A diagram of a generic data collection scheme where users collect data in some environment and then send the data through an optional mix network that can either be a geographical zone in their environment or a separate network. The data is eventually communicated to the data collector who may chose to use data obfuscation techniques to provide privacy to the users. The communication channel is represented by arrows.}
	\label{fig:dataCollection}
\end{figure}
Securing the communication channel from third parties that might want to intercept the data can be achieved using data encryption techniques. However, unless some additional steps are introduced, the entity collecting the data (a researcher or a company), from here on referred to as the \textit{data collector}, can link the data to its source. Giving pseudonyms to the participants can help mitigate this but it is still not completely safe. The data collector will still know that a certain batch of data belongs to a certain pseudonym which can be compromising depending on the content of the data. Even one piece of identifiable data will allow the data collector to know that all the data in that batch with the same pseudonym belongs to the same user. For this reason, mix networks were introduced in the data collection process. These networks mix the batches of data from each participant and send it to the data collector which then has no way to directly trace the source of a batch of data by essentially unlinking the data from all original network identifiers. 

Note that mixing alone is not sufficient to guarantee privacy, the data itself can still be used to identify a participant. For this, there has been a lot of research in data obfuscation which provides some level of anonymity (k-anonymity, l-diversity, t-closeness, and others) which should be used in addition to mix networks. Other privacy and security related requirements are outlined by Giannetsos et al. \cite{Giannetsos2014}, He et al. \cite{He2015a}, and Christin et al. \cite{Christin2016} however, these are out of the scope of this paper. These include privacy-preserving resilient incentive mechanisms and fairness (users should receive credits and rewards for their participation without associating themselves with the data or the tasks they contributed), communication integrity confidentiality and authentication (all entities should be authenticated and their communications should be protected from any alteration and disclosure to unauthorized parties), authorization and access control (participating users should act according to the policies specified by the sensing task), data-centric trust (Mechanisms must be in place to assess the trustworthiness and the validity of user submitted data), and accountability (entities should be held accountable for actions that could disrupt the system operation or harm users).

As we will see in section \ref{sec:oMPSrelatedWork} the concept of mix networks is not new, however, there are no studies that evaluate the performance of peer to peer opportunistic mixing using real world mobility data of the participants. In this paper, our goal is \textbf{not} to introduce a novel data mixing strategy but rather to focus specifically on evaluating the efficacy of mixing the data, as part of a slicing and mixing strategy, in a fully opportunistic way with the goal of achieving a uniform distribution of the data among all participants. We assume that participants are mobile and generate data by using sensors or answering surveys and that they regularly cross paths with at least one other participant in order to exchange data in a peer to peer manner. Furthermore, we require that all participants together form a connected graph with respect to who they meet. The mixing strategy we use is very basic so as to provide baseline results that can later be used to evaluate more complex strategies.

In section \ref{sec:oMPSrelatedWork} we present some related work on privacy techniques and data mixing in the context of participatory sensing, in section \ref{sec:oMPSmain} we describe a data mixing scheme, in section \ref{sec:oMPSresults} we verify the mixing scheme in a simulated environment. Finally, in section \ref{sec:oMPSconclusion} we summarize the results and provide design implications.

\section{Related Work}
\label{sec:oMPSrelatedWork}
There are several measures of anonymity for data content: $k$-anonymity is achieved when each data value occurs at least $k$ times, $l$-diversity is a stronger privacy indicator and ensures that there are at least $l$ well represented values for sensitive attributes based on entropy among other metrics, other stronger privacy indicators commonly used is $t$-closeness \cite{Sweeney2002}\cite{Machanavajjhala2006}\cite{Li2007b}. As we mentioned, these metrics are about the data content as a whole and are not relevant to this work which focuses on how to ensure privacy during data reporting at the source. 

If only aggregate information is needed from a set of data, privacy-preserving data aggregation schemes have been proposed in order to safely provide aggregate information such as value averages, minimums, and maximums about the underlying data using homomorphic encryption or other techniques \cite{Shi2010,Wei2013,Zhang2013a,Erfani2013,Zhang2016}. These aggregate values cannot be used by a researcher or other entity to train machine learning models like neural networks or SVMs which most often require the data to not be aggregated. It is advantageous to do the least amount of manipulation on the data in order to retain as much utility as possible.

A common practice when performing studies is to use pseudonyms for the participants while keeping the data at its pure form. However it has been shown overwhelmingly that this does not necessarily guarantee the privacy of the participants \cite{Scipioni2010,Christin2011a,Shin2015}. Other approaches use mix networks or mix zones to reassign pseudonyms to the participants or to mix data. These methods have been shown to be an effective way to protect the participant's identity by decoupling the data from the user who collected it \cite{Chaum1981,Cornelius2008,Neff2001}. Mix networks are well studied and quite robust at what they are designed to do, which is to shuffle the batches of data (i.e. permuting the order of the batches with respect to the pseudonyms). The limitation of this approach is that the participants who generate the data often have to trust the mix network and additionally, for many mix network designs, individual data entries remain in the original batch so that when the identity of the participant is discovered for one piece of data from a batch then the rest of the batch can be assumed to belong to that same participant. Mix zones are fixed in space and require that participants enter these zones to satisfy certain privacy aspects like k-anonymity by guaranteeing that the data is mixed among $k$ participants making them unsuitable for opportunistic settings.

A novel approach to data privacy is \textit{slicing and mixing}. First developed for wireless sensor networks, it partitions the data horizontally and then mixes it before aggregating values (SMART) \cite{He2007}. It was adapted for privacy in data publishing for human-generated data by partitioning the data both vertically and horizontally where in the former, care is taken to group highly correlated attributes together. Then these \textit{slices} are permuted in order to break the linking between different columns \cite{Li2012a}. Many works have extended slicing to be used in participatory sensing scenarios for privacy-preserving data aggregation\cite{Shi2010} but only a few have looked into how the mixing might perform in real world environment with real people (ex. Qiu et al. \cite{Qiu2013} use taxi traces to simulate participants) and none to our knowledge do an analysis of the effectiveness of mixing when it comes to opportunistic peer to peer (P2P) mixing scenarios. Christin et al. \cite{Christin2011b,Christin2013b} also used P2P techniques to obfuscate location information however, their evaluation does not generalize to other types of data.

\section{Privacy-conscious Data Shuffling}
\label{sec:oMPSmain}
From here on we will refer to the participants (users in Figure \ref{fig:dataCollection}) who are generating the data as \textit{sensor nodes} or just \textit{nodes}. Our simple opportunistic shuffling scheme consists of the following steps: Once a node $i$ meets another node $j$, they randomly select some of their data in order to swap it with each other. The nodes follow their regular mobility patterns and are exchanging data with each other when they come into direct communication range of each other until certain stopping criteria relating to the state of the shuffle are met. As we mentioned earlier, we require that the nodes form a connected graph so that the entirety of the data can be uniformly shuffled. If there are any disconnected subsets of nodes the data will have no way to be communicated between those subsets, only within them.

\subsection{Data exchange}
\label{sec:oMPSdataExchange}
When two nodes come within communication range of each other, each node randomly selects \textbf{half} of the data they have in their possession, $M$, to exchange with the other. 
This value can be adjusted individually on each node to adjust their data exposure and either reduce or increase the potential amount of personal data that they might share at each transaction. 
The specifics of this adjustment are not explored in this paper and we keep the amount of data that each node exchanges at $\frac{M}{2}$ as it is optimal for reaching a uniform distribution of data in the least number of shuffles. 
This fact can be easily verified by looking at the number of ways there are to choose $k$ data from $M$ given by the binomial coefficient which can be calculated using the equation below:
\begin{equation}
\begin{pmatrix} k\\M\end{pmatrix} = \frac{M!}{k!\left(M-k\right)!}
\label{eqn:binomialCoef}
\end{equation}
Each shuffle becomes more random as the binomial coefficient increases in value. If we set $k$ to be some fraction $x\in\left[0,1\right]$ of the total data $M$, $k=xM$, then we can find that the value of $x$ which maximizes the result in the equation above is $\frac{1}{2}$.

\subsection{Stopping criteria}
\label{sec:oMPSstopCriteria}
There are two different stopping criteria that can be used to signify that the data has been sufficiently shuffled (uniformly distributed) and that it is safe to send it to the data collector. 

The first one is based on each nodes perception of how well the data is mixed. Each node can keep track of the data that they come into contact with and measure the probability that they encounter some specific piece of data. Since the data may be encrypted, the nodes must keep track of the encrypted data or a shorter hashed version of the encrypted data which can come paired up with the encrypted data. Once the probability is close to being uniform across all data, then they can stop the shuffling process since this indicates a near uniform mix. This might work well when there are not many nodes, but as the number of nodes increases, the time it takes to verify the uniformity of the mix also increases. 

The second set of stopping criteria is based on the properties of the graph like closeness centrality. If, in addition to data, the nodes exchanged information about their graph connections (nodes that they have previously encountered) or there is prior knowledge about the graph, then they can estimate the number of exchanges that they need to perform before the data is near uniformly mixed.

\paragraph{Closeness centrality to estimate stopping criteria.}
\label{sec:oMPScentrality}
Closeness centrality is a measure of the degree to which an individual node is near all other nodes in the network. In order for each node to calculate its closeness centrality it needs to know its distance to all other nodes. This is trivial when there is global knowledge about the graph, however, it may not always be the case, especially when there is no trusted party to provide this information. When there is no prior graph knowledge, each node $i$ needs to communicate its personal adjacency matrix $A^i$ in addition to the data at each exchange. $A^i$ should initially indicate which nodes are directly connected (one hop) along with the edge weight as it is calculated by the node (in this case, edge weight is equivalent to the number of times that the node has encountered each of its one hop neighbors). 
Then the node can update $A^i$ by combining all the personal adjacency matrices it has acquired ($A^j,A^k,$ etc.) from other nodes. To update $A^i$ when the node receives another node's personal adjacency matrix $A^j$ we perform the following operation:
\begin{algorithm}
	\DontPrintSemicolon
	\SetAlgoNoLine
	\KwData{$A^i$, $A^j$}
	\KwResult{updated $A^i$}
	\If{$A^{i}_{k,l} = \emptyset$ and $A^{j}_{k,l} \neq \emptyset$}{
		$A^{i}_{k,l} \leftarrow A^{j}_{k,l}$\;
	}
	\If{$A^{i}_{k,l} \neq \emptyset$ and $A^{j}_{k,l} \neq \emptyset$}{
		$A^{i}_{k,l} \leftarrow min\left(A^{i}_{k,l},A^{j}_{k,l}\right)$\;
	}
	\caption{Procedure to combine $A^i$ with $A^j$}
	\label{algo:updateAi}
\end{algorithm}

\begin{figure}[H]
	\centering
	\includegraphics[width=0.6\textwidth]{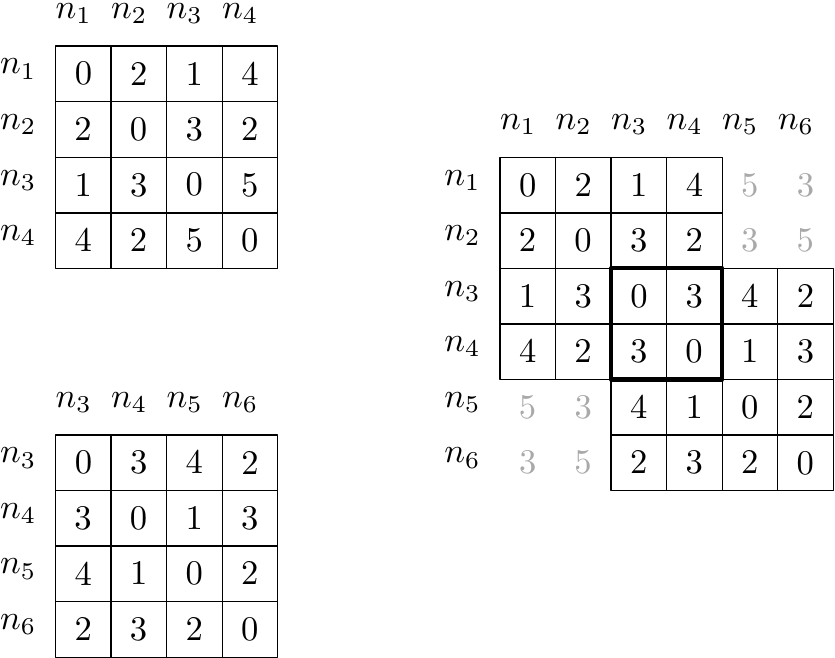}
	\caption{On the left, two personal adjacency matrices. On the right, the combined outcome where overlapping weights are assigned the minimum of the two values. Grey values indicate values calculated by the shortest path algorithm that have not been directly measured otherwise.}
	\label{fig:combineAdjacency}
\end{figure}

Finally, performing a shortest path algorithm such as the Floyd-Warshall or Dijkstra algorithm can reduce the redundancies and update the paths in $A^i$. 
The process is illustrated in figure \ref{fig:combineAdjacency}

The closeness centrality can then be calculated for each node and by each node using the information in their respective adjacency matrix.

The stopping criteria for the number of exchanges necessary to sufficiently mix the data can then be estimated from the adjacency matrix and closeness centrality information using empirical data which we will show in section \ref{sec:oMPSresults}.

It is important to note that if the graph of the nodes is known to everyone, encrypting the communication channel becomes even more vital for the protection of the security and privacy of nodes against malicious nodes. 

\section{Experiment and Results}
\label{sec:oMPSresults}
\subsection{Experimental setup}
\label{sec:oMPSexpSetup}
To verify our data mixing scheme, we perform simulations using artificial parameters as well as simulations using real mobility data from the MDC dataset. The data mixing occurs in shuffling rounds that consist of either a group of markov chain state transitions (representing data exchanges) based on the transition matrix or a full day (24 hours) of proximity events in the real mobility data simulations.

At each time $t$, a node $i$ will exchange data with a node $j$ either with a probability based on the $A_{i,j}$ element of the transition matrix $A$ for the artificial parameter simulations or based on the proximity of the two nodes in the real mobility data simulations. At each shuffle we take note of what data each node has. 

In order to get representative probability distributions of the data, we run $30000$ trials of the simulation with each of the parameter sets (a parameter set consists of the following: the number of nodes, data size per node, and the transition matrix or proximity events). This number was selected because it gives us a confidence level of $99\%$ based on the equation $n\geq\frac{log\left(a\right)}{log\left(1-p\right)}$ to calculate the number of trials necessary given the probability of the occurrence of an event $p$ and the confidence level $1-a$ that we require. In our case we seek to be confident of events that occur with a probability of at least $p=0.00015$, that is to say that our probability distributions in our results will have a granularity of $0.00015$. We choose $1-a=0.99$ which is equivalent to being $99\%$ confident in our results.

For our experiments, the total number of nodes, $N$, and the amount of data items, $M$, that they start with is selected as described in section \ref{sec:oMPSparamSets}. To keep track on how the data flows throughout the network we make sure that each node's initial data is uniquely identifiable by labeling them with integers. For example, if we set the number of data to $6$, then node $1$'s data will consist of the numbers from $1$ to $6$, node $2$'s data will consist of the numbers from $7$ to $12$, etc.. In this way we can easily check how uniformly the data has been distributed by evaluating the probability distribution of each number being in any specific node at the end of the shuffle.

\subsubsection{Artificial Parameter Simulations Setup}
Initially, we performed simulated experiments with artificial parameters to illustrate and validate the data shuffling procedure. Node mobility is artificially simulated by using three different Markov models where each one is defined by a transition matrix $A$. The three models consist of a best case scenario transition matrix (equivalent to a group of co-workers or students enrolled in the same course), an intermediate case scenario (equivalent to shift-based co-workers), and a worst case scenario transition matrix (equivalent to otherwise unrelated commuters crossing paths on their way to their individual workplaces). The specifics of the transition matrices are described in section \ref{sec:oMPSparamSets}.

\subsubsection{Real Mobility Data (MDC) Simulations Setup}
We use real user mobility traces from the data of the Mobile Data Challenge (MDC) which include GPS traces of real mobile users \cite{Kiukkonen2010,Laurila2012}. The data we used included $191$ users and spanned over a year of data for some of the users. To normalize and be able to compare with the artificial cases, we define one shuffling round as a single day and we analyze up to $100$ contiguous days (i.e. $100$ shuffling rounds) of GPS traces for each trial. 

For some users, we might have less than $100$ days of data, when we reach the end of the data without having completed the $100$ shuffling rounds we cycle from the beginning until we reach the desired number. For example, if a user set only has $50$ days worth of data, we will go though his GPS data twice to complete the $100$ day trial.

In most cases we have more than $100$ days of data for each user set. In this case, since we limit our simulations to $100$ shuffling rounds consisting of $100$ contiguous days, we make sure that we select $100$ contiguous days when the user set is sufficiently active based on two criteria in order of priority: the median of the number of proximity events between all pairs in the user set, and the total number of proximity events.

We assume an exchange of data between two users can be performed under the following conditions: 
\begin{itemize}
	\item They are within $50$ meters or less of each other. We call this a \textit{proximity event} since they are within direct communication range of each other.
	\item They have not exchanged data between each other in the past 30 minutes.
\end{itemize}
In these experiments we do not consider the bandwidth or throughput of the data transmission and assume that it can be instantaneously exchanged when two users are within communication range.

For each of the $30000$ trials for the MDC data simulations, a random subset of $N$ users was selected from the $191$ in such a way that they formed a connected graph with a minimum edge weight of $10$ (in this case, edge weight indicates the number of exchanges between two users during the entire duration of the study). With this random selection, when we use $N=10$ the average number of hops between the two most remote users was $9$ and the median number of hops between any two users was $3$ (which resembles a line topology). The user set was unique in each trial of our simulations, i.e. no two trials had the same set of $10$ random users.

We ran an additional simulation with the MDC dataset in which instead of selecting $N$ random users, we selected $N$ users that formed a \textit{clique} (i.e. a fully connected topology with maximum distance of one hop between any two users). Again, we limited the edge weight to be equal to or above $10$. During our experiments we discovered that there were not enough cliques of size $\geq N$ in the dataset to justify doing $30000$ trials. The total number of maximal cliques (cliques that are not subsets of larger cliques) in the dataset is $890$ and the number of cliques with size of at least $N$ is often much smaller than that. It is redundant to perform more trials than there are number of cases because this means that the same case will need to be repeated several times to reach the desired amount of trials. However, to get meaningful statistics it was necessary to do a much larger number of trials than there were number of cliques. To remedy this, we relaxed the requirement for the cliques and allowed ourselves to combine cliques to form a set of $N$ users. The exact procedure by which we combined the cliques is described in Algorithm \ref{algo:cliqueCombination}. The \textbf{if} statement on lines 4-6 is optional and serves to limit the minimum size of the cliques that form the user set thus ensuring a better connected user set. 
\begin{algorithm}
	\DontPrintSemicolon
	\SetAlgoNoLine
	\KwData{The user cliques}
	\KwResult{A well connected user set}
	$userSet\leftarrow \emptyset$\;
	\While {$size(userSet)<N$}{
		$cliq \leftarrow random\,clique $\;
		\If{$ size(cliq) < 0.5\left(N-size\left(userSet\right)\right)$}{
			go to line 3\;
		}
		$userSet \leftarrow userSet \cup cliq$\;
		\If{size($userSet)>N$}{
			$userSet\leftarrow $ select $N$ users $\in userSet$\;
		}
	}
	\If{$userSet$ not connected }{
		go to line 1\;
	}
	\caption{Procedure to combine cliques}
	\label{algo:cliqueCombination}
\end{algorithm}
This algorithm allowed us to generate much more than $30000$ different user sets as evidenced by the results in our simulations where for $N=10$ there were no two trials with the same user set in all of the $30000$ trials. Furthermore, $N=10$ resulted in user sets with the average number of hops between the two most remote users at $4$ and a median number of hops between any two users of $1$.

\subsubsection{Parameter Sets}
\label{sec:oMPSparamSets}
\begin{table}
	\begin{center} 
		\begin{tabular}{lp{0.4\textwidth}c}
			\toprule
			\textbf{Sim. type} & \textbf{Transition Matrix} & \textbf{\{\# of Nodes, \# of Data\}}\\
			\midrule
			best case & fully connected topology w/ prob of no transaction $0$ and prob of transation with each of the other nodes $\frac{1}{N-1}$ & \{10,6\} \{30,6\} \{10,20\} \\
			intermediate case & line topology w/ prob of no transaction $0.5$ for edge nodes and $0$ for all other nodes & \{10,6\} \{30,6\} \{10,20\} \\
			worst case & line topology w/ prob of no transaction $0.8$ for edge nodes and $0.6$ for all other nodes & \{10,6\} \{30,6\} \{10,20\} \\
			MDC data random & GPS traces of a random selection of users & \{10,6\} \\
			MDC data cliques& GPS traces of cliques of users & \{10,6\} \\
			\bottomrule
		\end{tabular}
	\end{center}
	\caption{Simulations performed showing total nodes and total amount of data per node for each simulation. There are 11 simulations in total.}
	\label{table:sim}
\end{table}

The number of nodes and number of data items for the MDC data simulations was selected after the artificially simulated cases where we verified that the number of data items did not significantly affect the number of shuffles needed since we always exchange half of a node's total data (as per the protocol discussed in section \ref{sec:oMPSdataExchange}). We chose to simulate only 2 representative cases with the MDC dataset: choosing a connected set of random users, or choosing users that form cliques in the adjacency matrix. Other cases would be redundant since we already show the effects of changing the number of nodes and data items with the artificial parameter simulations. All simulations are summarized in Table \ref{table:sim}.

\subsection{Performance Criteria}
\label{sec:oMPSperfCriteria}
The performance criteria that we mainly use is the Kolmogorov-Smirnov test with a uniform distribution of $\frac{1}{N}$ as the reference distribution. With this test, we measure the absolute error between the distribution of the data in our experiment and the ideal uniform distribution. As a result of our experimental setup we are able to perform this test after every shuffling round in our experiment allowing us to see the exact number of shuffles needed to achieve a near uniform mix.

For illustrative purposes we first take a look at the probability of holding a specific data item for each node at each shuffling round.

\subsection{Results}
\subsubsection{Results Using Artificial Parameters}
\begin{figure}[H]
	\centering
	\begin{subfigure}[b]{0.48\textwidth}
		\centering
		\includegraphics[trim=50 190 50 190, clip, width=\textwidth]{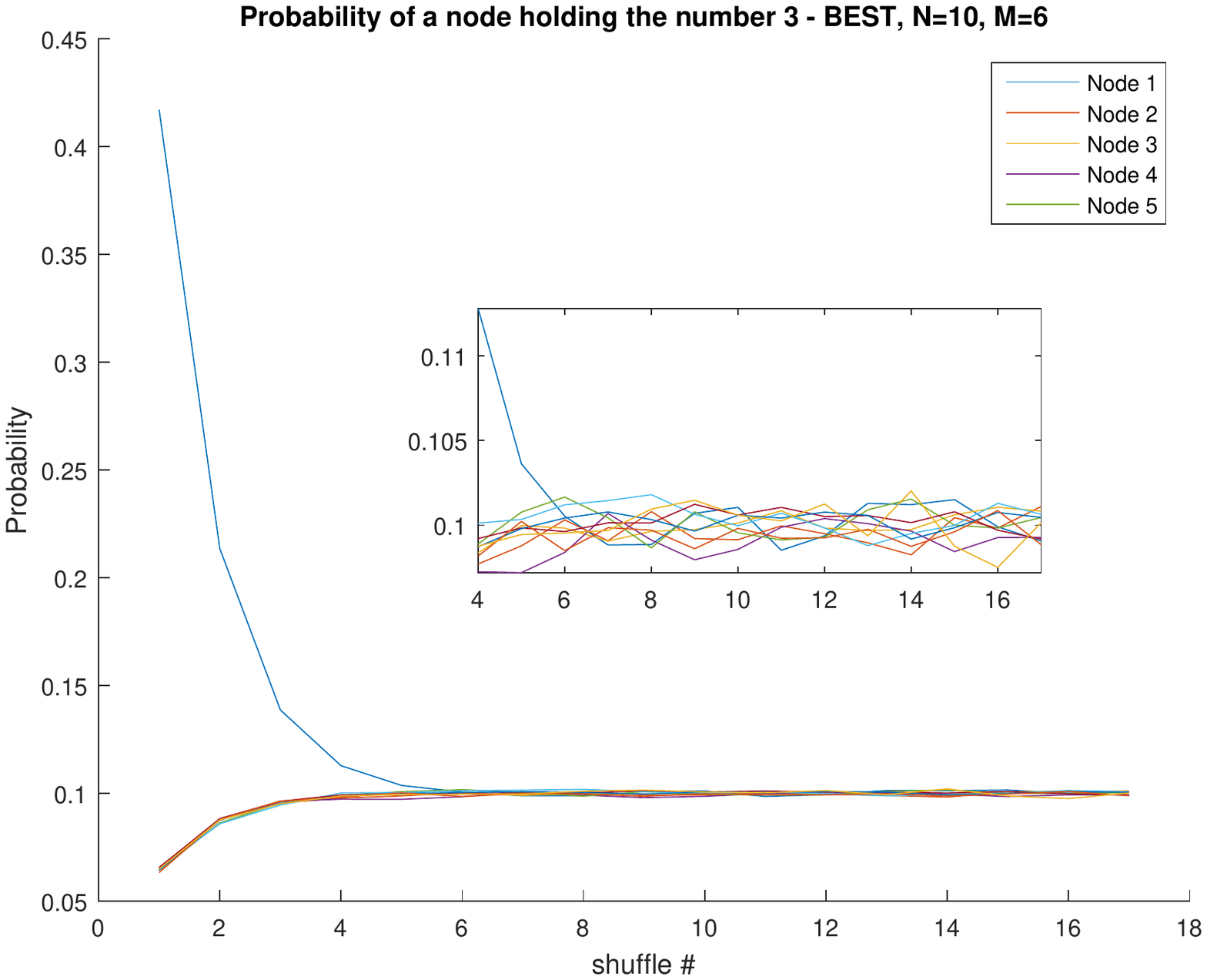}
		\caption{The probability distribution of the number $3$ being at different nodes at different number of shuffles (subplot is of a magnified region)}
		\label{fig:probOf3_best}
	\end{subfigure}
	\begin{subfigure}[b]{0.48\textwidth}
		\includegraphics[trim=50 190 50 190, clip, width=\textwidth]{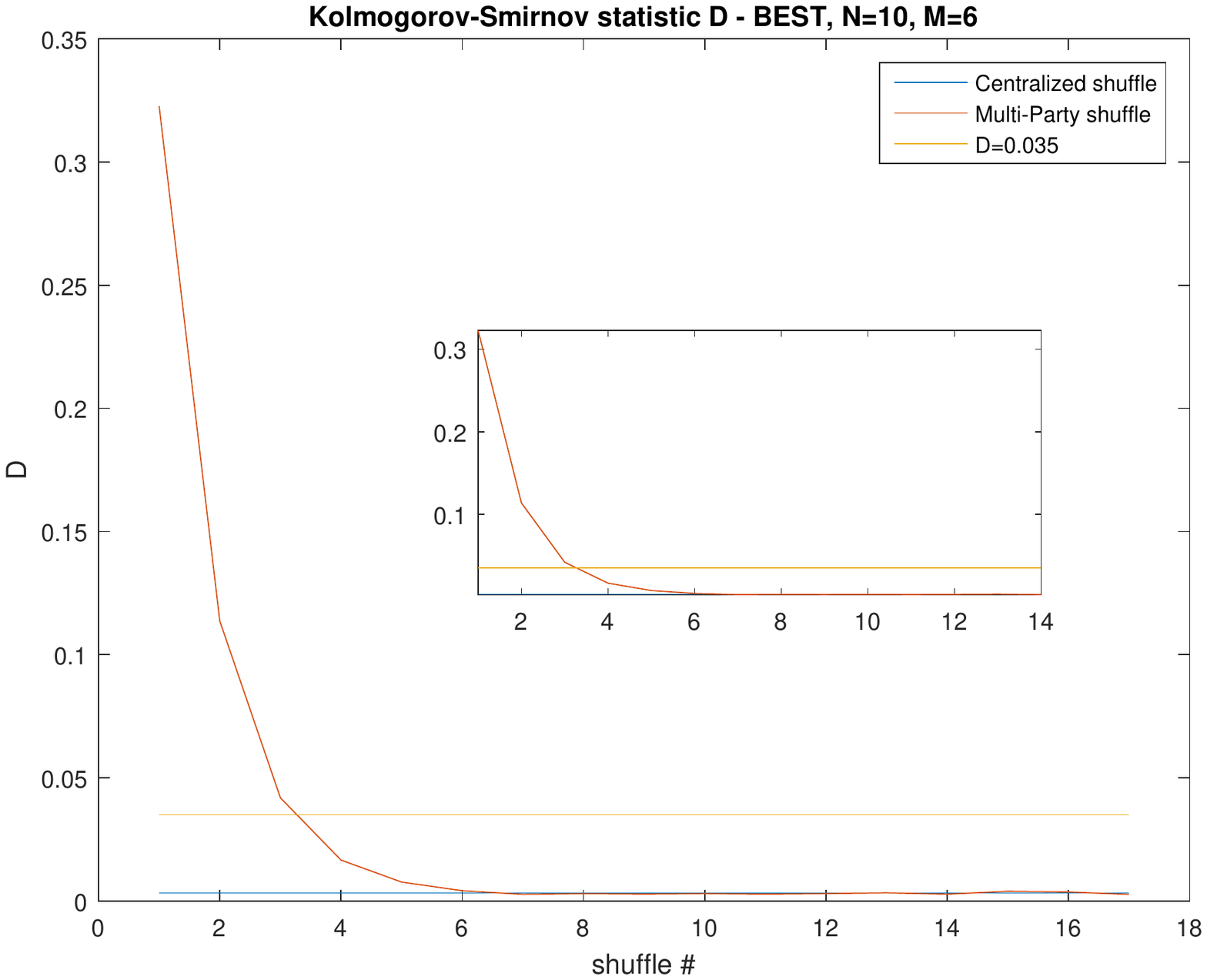}
		\caption{The results of the Kolmogorov-Smirnov test at different number of shuffles(subplot is of a magnified region)}
		\label{fig:ksStatistic_best}
	\end{subfigure}
	\caption{Results for the best case scenario for $N=10$ and $M=6$}
	\label{fig:results_best}
\end{figure}

\begin{figure}[H]
	\centering
	\begin{subfigure}[b]{0.48\textwidth}
		\centering
		\includegraphics[trim=50 190 50 190, clip, width=\textwidth]{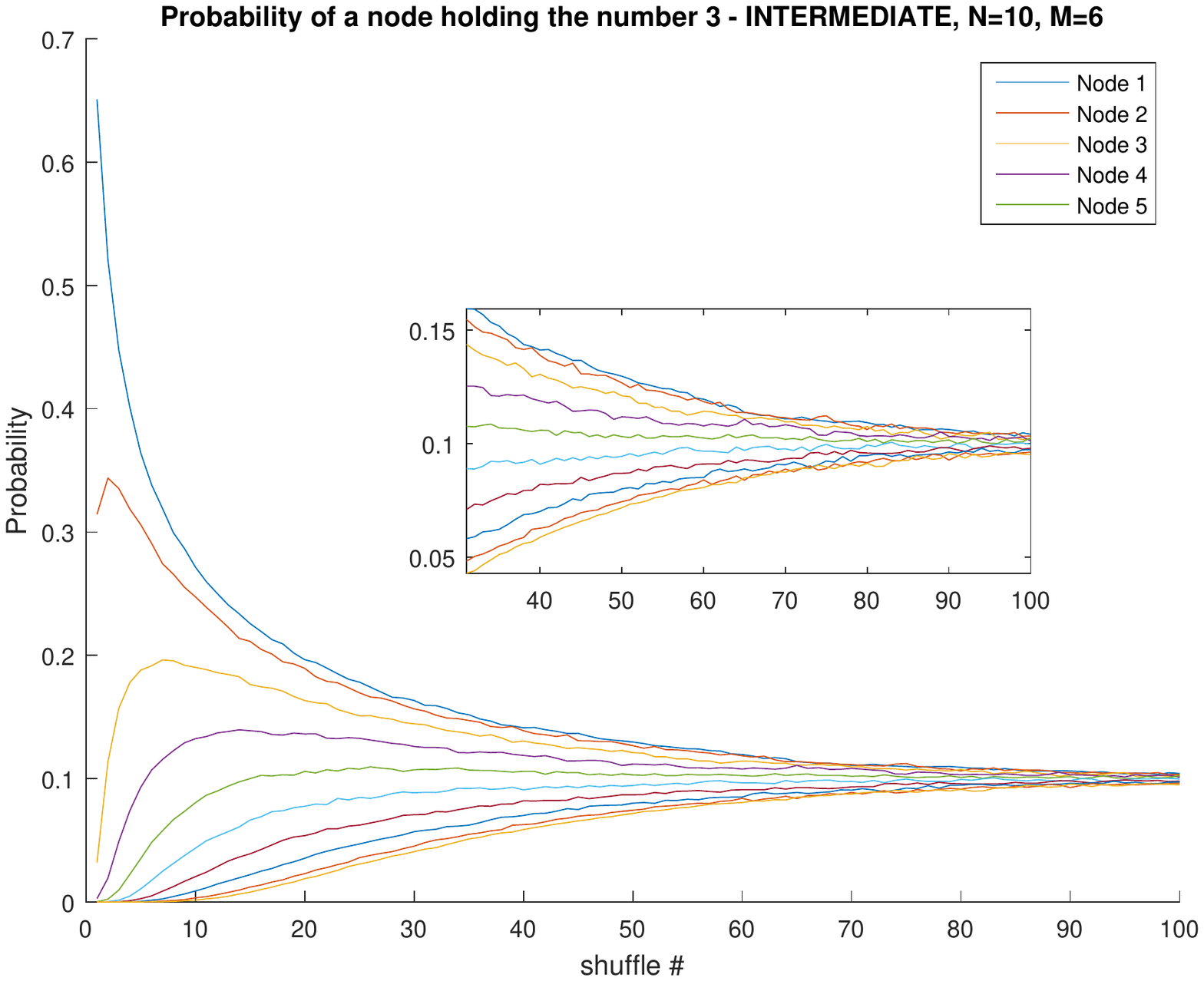}
		\caption{The probability distribution of the number $3$ being at different nodes at different number of shuffles (subplot is of a magnified region)}
		\label{fig:probOf3_med}
	\end{subfigure}
	\begin{subfigure}[b]{0.48\textwidth}
		\includegraphics[trim=50 190 50 190, clip, width=\textwidth]{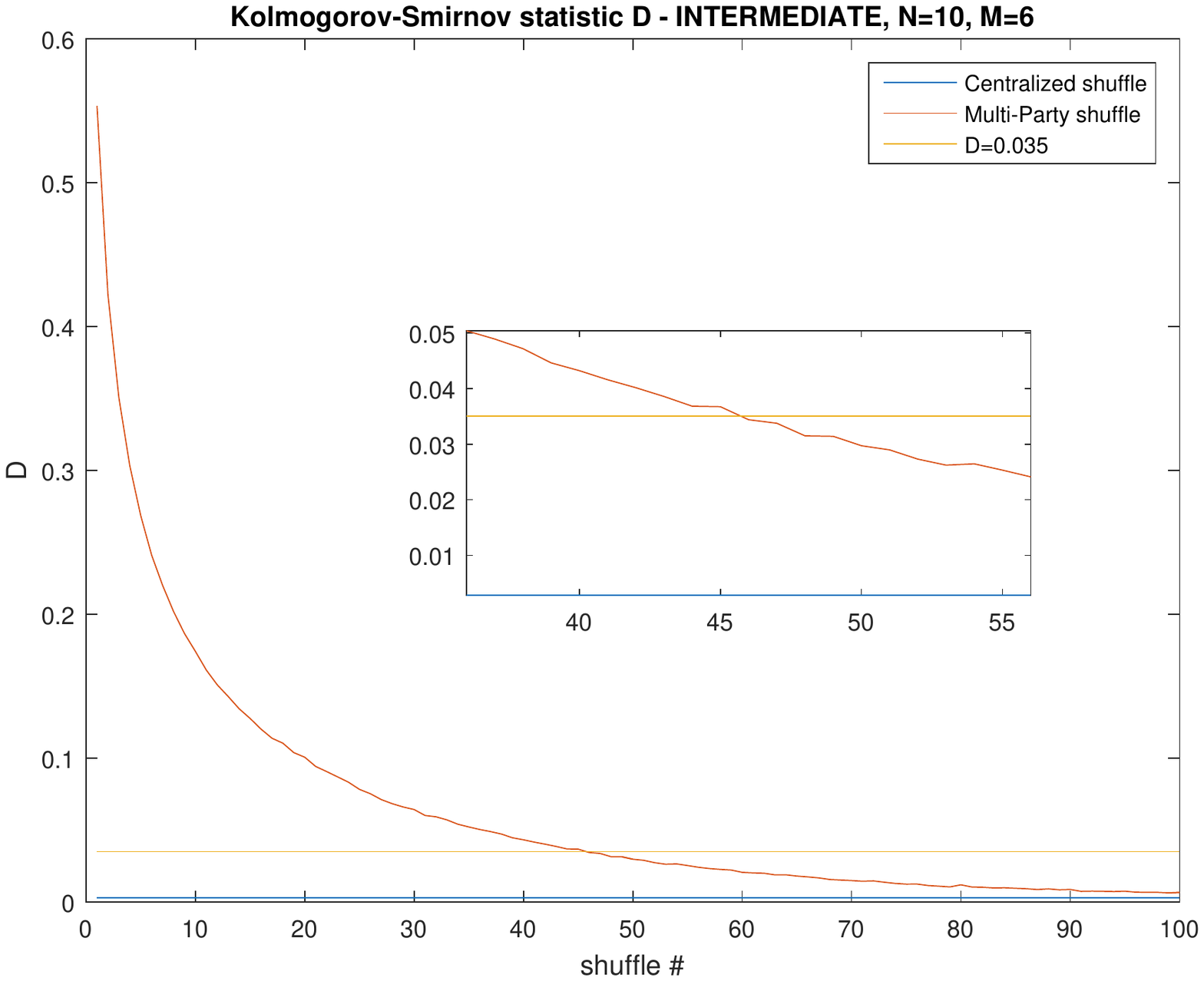}
		\caption{The results of the Kolmogorov-Smirnov test at different number of shuffles(subplot is of a magnified region)}
		\label{fig:ksStatistic_med}
	\end{subfigure}
	\caption{Results for the intermediate case scenario for $N=10$ and $M=6$}
	\label{fig:results_med}
\end{figure}

\begin{figure}[H]
	\centering
	\begin{subfigure}[b]{0.48\textwidth}
		\centering
		\includegraphics[trim=50 190 50 190, clip, width=\textwidth]{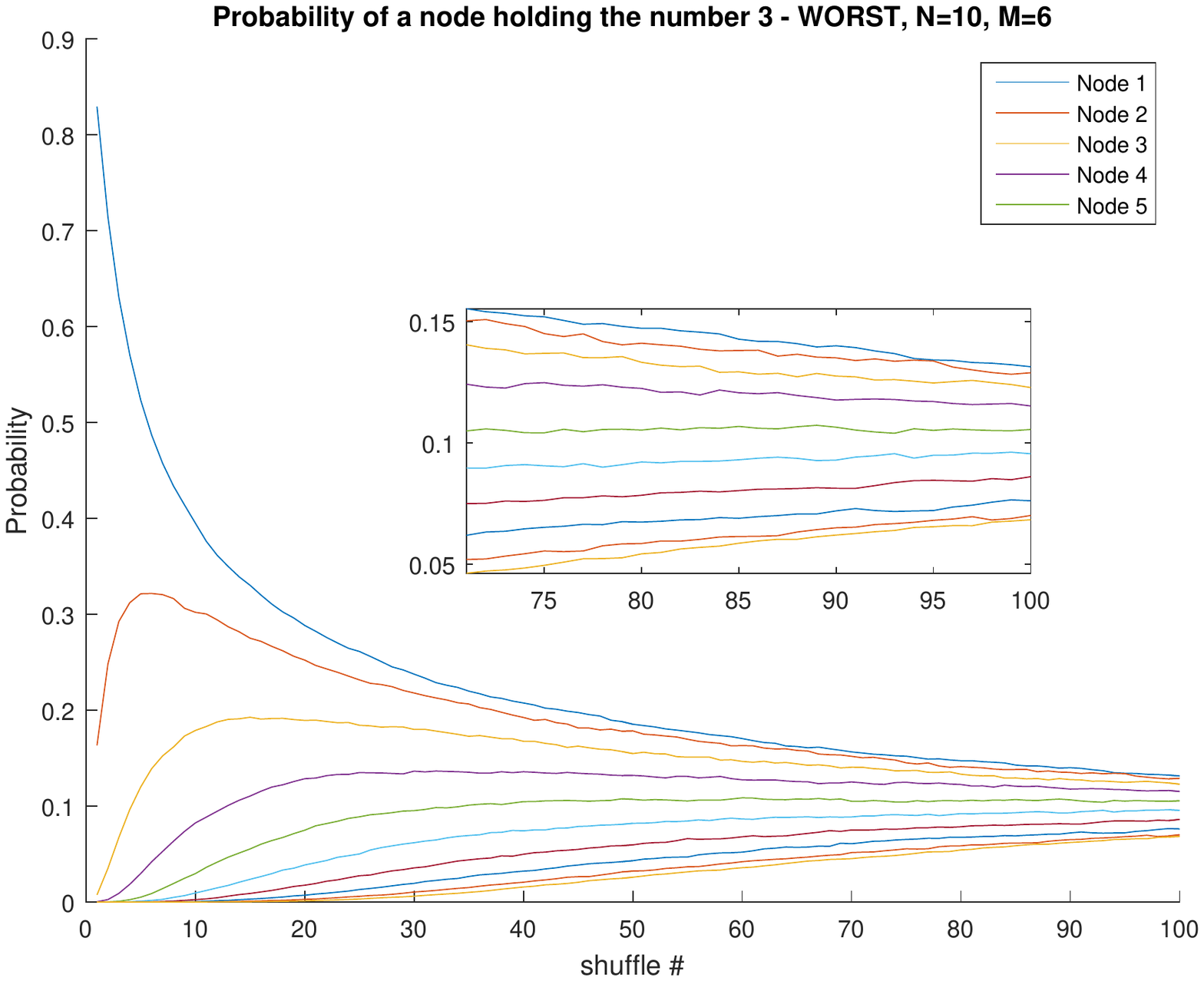}
		\caption{The probability distribution of the number $3$ being at different nodes at different number of shuffles (subplot is of a magnified region)}
		\label{fig:probOf3_worst}
	\end{subfigure}
	\begin{subfigure}[b]{0.48\textwidth}
		\includegraphics[trim=50 190 50 190, clip, width=\textwidth]{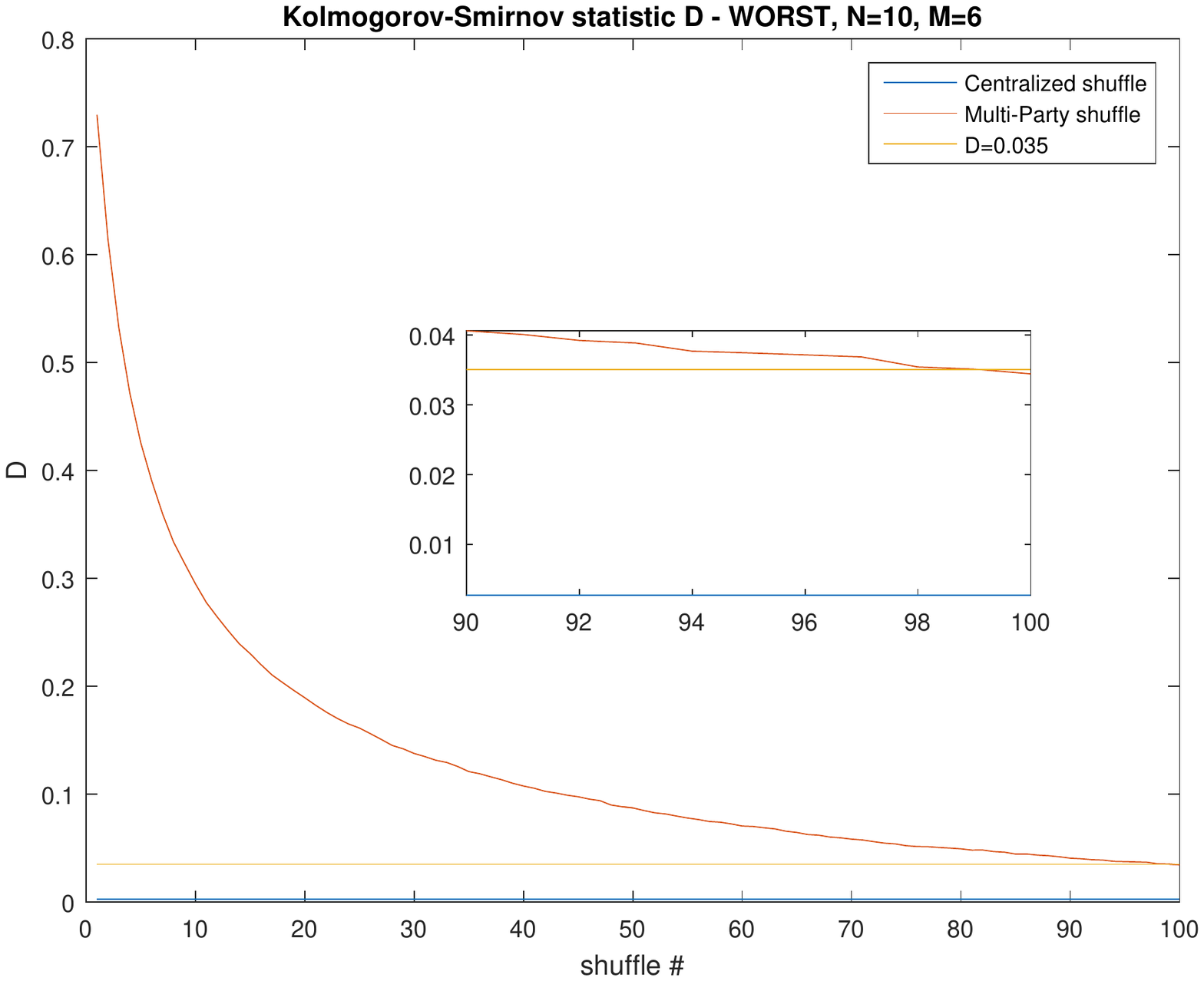}
		\caption{The results of the Kolmogorov-Smirnov test at different number of shuffles(subplot is of a magnified region)}
		\label{fig:ksStatistic_worst}
	\end{subfigure}
	\caption{Results for the worst case scenario for $N=10$ and $M=6$}
	\label{fig:results_worst}
\end{figure}

Intuitively, the more shuffles we do then the more uniform the distribution of the data should be. This intuition is verified in figure \ref{fig:probOf3_best} where we clearly see that the probability of holding a specific data item (as an example we use the data item with number $3$) approaches an ideal probability with amplitude $\frac{1}{N}$ as the number of shuffles increases, where $N$ is the number of nodes. Since node $1$ is the initial holder of the data item with number $3$, it starts with the highest probability in the initial stages and as it shares data with all the other nodes the probability evens out.
The same is true for the intermediate and worst case of the line topology as we can see in figures \ref{fig:probOf3_med} and \ref{fig:probOf3_worst}, although, in this case it takes more than $40$ shuffling rounds to reach the same ideal probability for each of the cases. Similarly to the best case, we notice that node $1$ starts out with higher probability of holding the data with number $3$, and then we notice a sharp increase on the probability of node $2$ holding it since it is the only node that is connected to node $1$ (recall that intermediate and worst case scenarios have the line topology of nodes). 

\subsubsection{Results using MDC Dataset}
\begin{figure}[H]
	\centering
	\begin{subfigure}[b]{0.48\textwidth}
		\centering
		\includegraphics[trim=50 190 50 190, clip, width=\textwidth]{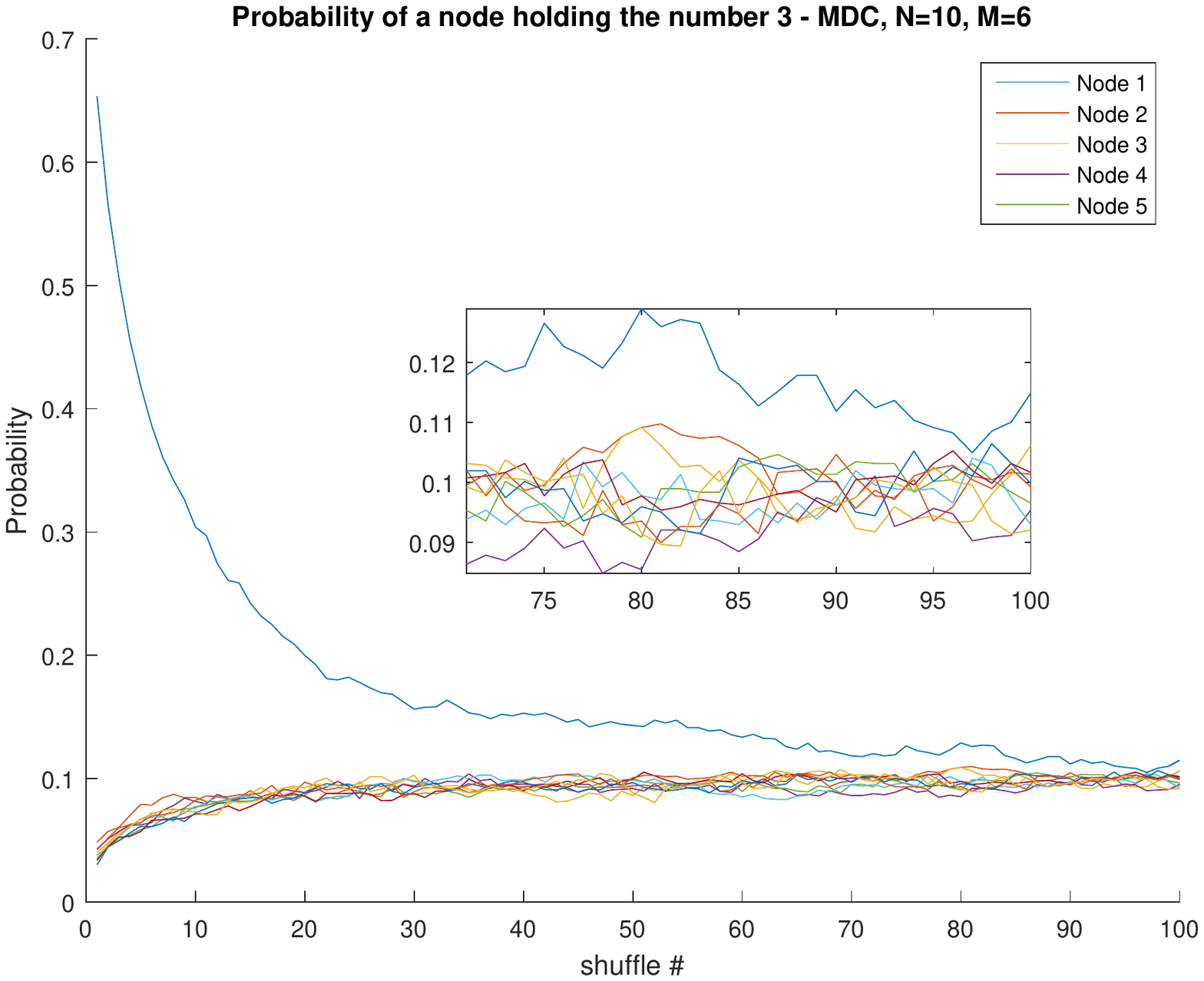}
		\caption{The probability distribution of the number $3$ being at different nodes at different number of shuffles (subplot is of a magnified region)}
		\label{fig:probOf3_mdc}
	\end{subfigure}
	\begin{subfigure}[b]{0.48\textwidth}
		\includegraphics[trim=50 190 50 190, clip, width=\textwidth]{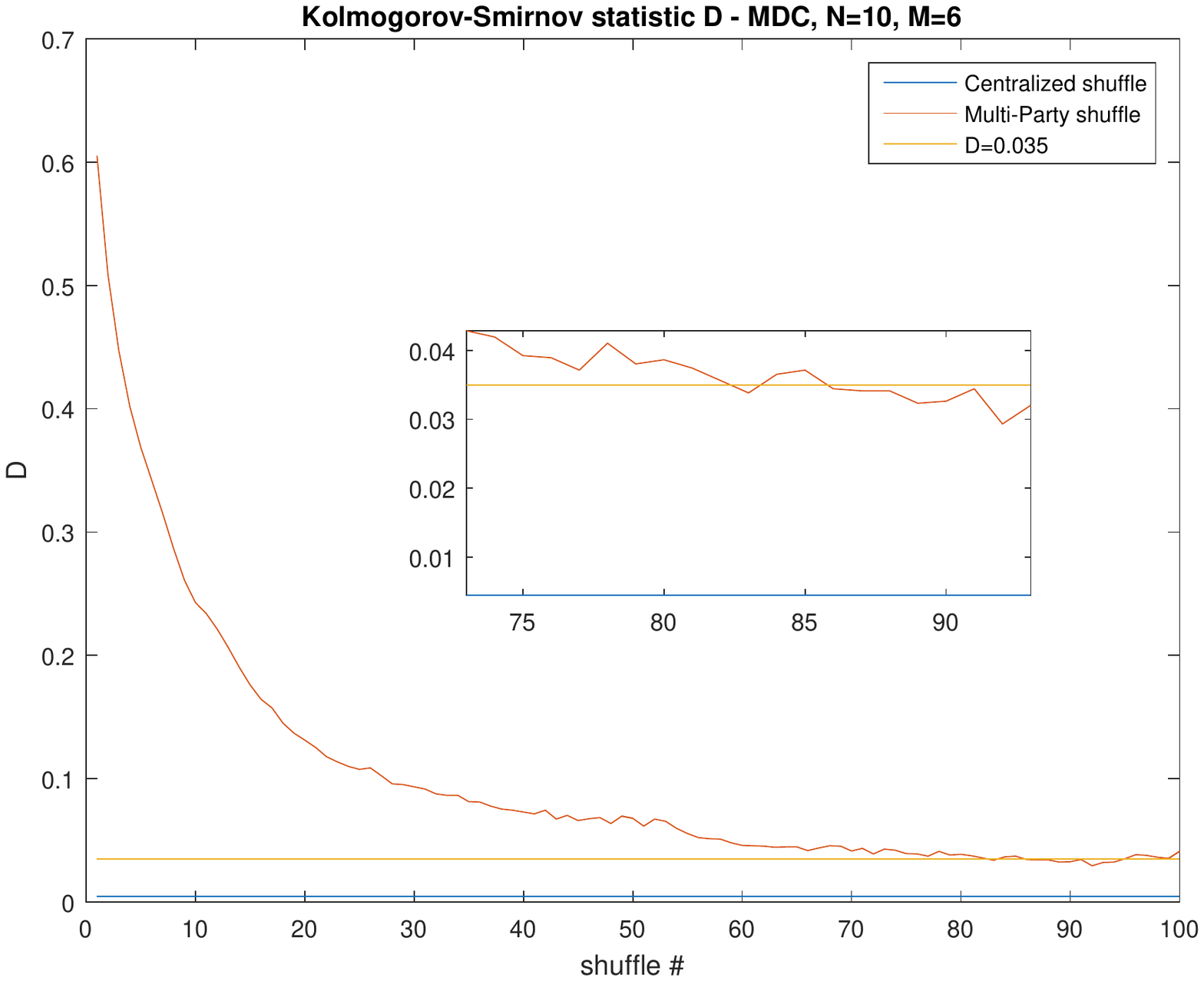}
		\caption{The results of the Kolmogorov-Smirnov test at different number of shuffles(subplot is of a magnified region)}
		\label{fig:ksStatistic_mdc}
	\end{subfigure}
	\caption{Results for the MDC data with random user selection for $N=10$ and $M=6$}
	\label{fig:results_mdc}
\end{figure}

\begin{figure}[H]
	\centering
	\begin{subfigure}[b]{0.48\textwidth}
		\centering
		\includegraphics[trim=50 190 50 190, clip, width=\textwidth]{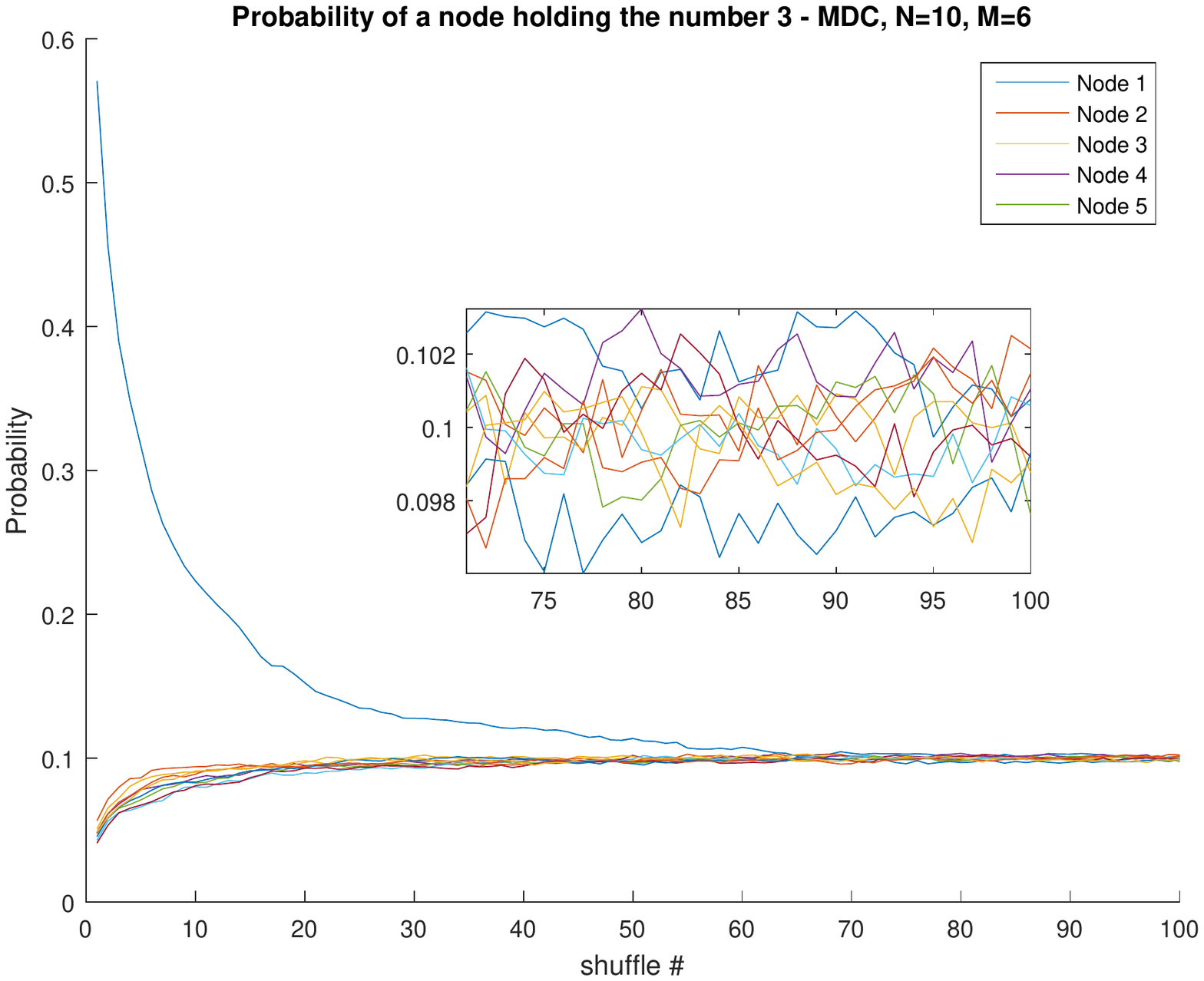}
		\caption{The probability distribution of the number $3$ being at different nodes at different number of shuffles (subplot is of a magnified region)}
		\label{fig:probOf3_mdccl}
	\end{subfigure}
	\begin{subfigure}[b]{0.48\textwidth}
		\includegraphics[trim=50 190 50 190, clip, width=\textwidth]{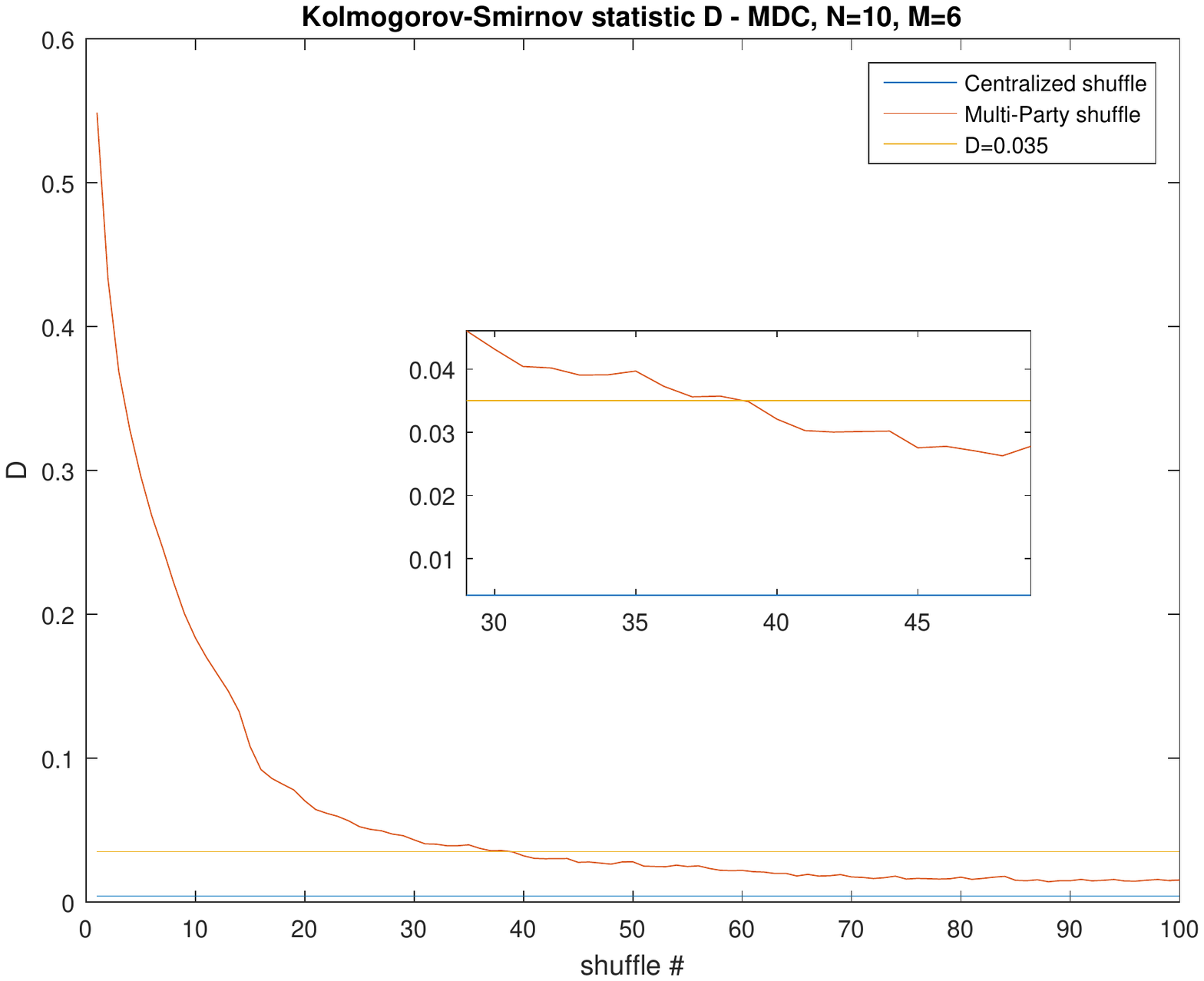}
		\caption{The results of the Kolmogorov-Smirnov test at different number of shuffles(subplot is of a magnified region)}
		\label{fig:ksStatistic_mdccl}
	\end{subfigure}
	\caption{Results for the MDC data with clique user selection for $N=10$ and $M=6$}
	\label{fig:results_mdccl}
\end{figure}

In figures \ref{fig:probOf3_mdc} and \ref{fig:probOf3_mdccl} we see the results of the MDC dataset simulations. For the MDC simulation with the random selection of users, although the connectivity resembles that of line topology, we cannot see it in figure \ref{fig:probOf3_mdc} (like in figures \ref{fig:probOf3_med} and \ref{fig:probOf3_worst}) because the users are not ideally ordered at each trial to reveal the same pattern as the artificial parameter simulation with line topology (recall that they were randomly selected).

\subsubsection{Kolmogorov-Smirnov test of the experiment}
As we can see in figures \ref{fig:ksStatistic_best}, \ref{fig:ksStatistic_med}, \ref{fig:ksStatistic_worst}, \ref{fig:ksStatistic_mdc}, and \ref{fig:ksStatistic_mdccl}, the error decreases as the number of shuffles increases. For our experimental setup, $4$ shuffles is sufficient to adequately shuffle the data in the best case scenario while more than $60$ shuffling rounds are needed in order to reach a similar distribution of the data for the worst case scenario. The MDC dataset simulation with random user selection is comparable to the worst case scenario while with clique based user selection it is better than the intermediate case but worse than the best case scenario.
\begin{table}
	\begin{center} 
		\begin{tabular}{lc}
			\toprule
			\textbf{Simulation type} & \textbf{\# of shuffles for $D<0.035$}\\
			\midrule
			Best case & $4$ \\
			Intermediate case & $46$ \\
			Worst case  & $100$ \\
			MDC data random & $85$ \\
			MDC data cliques & $40$ \\
			\bottomrule
		\end{tabular}
	\end{center}
	\caption{Summary of Kolmogorov-Smirnov test results for $N=10$ $M=6$.}
	\label{table:simSummary}
\end{table}

\subsubsection{Effects of varying number of nodes and number of data items}
\begin{figure}[H]
	\centering
	\begin{subfigure}[b]{0.48\textwidth}
		\centering
		\includegraphics[trim=50 190 50 190, clip, width=\textwidth]{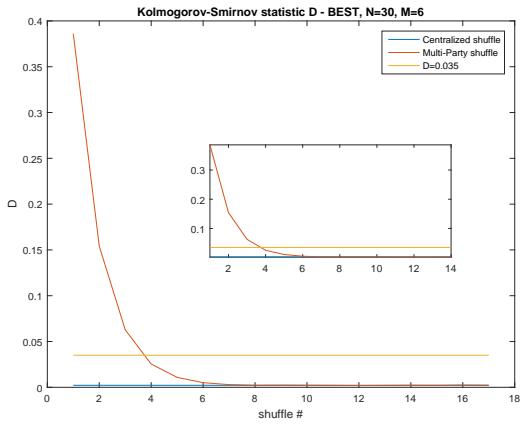}
		\caption{$N=30$, $M=6$}
		\label{fig:ksStatistic_best1}
	\end{subfigure}
	\begin{subfigure}[b]{0.48\textwidth}
		\includegraphics[trim=50 190 50 190, clip, width=\textwidth]{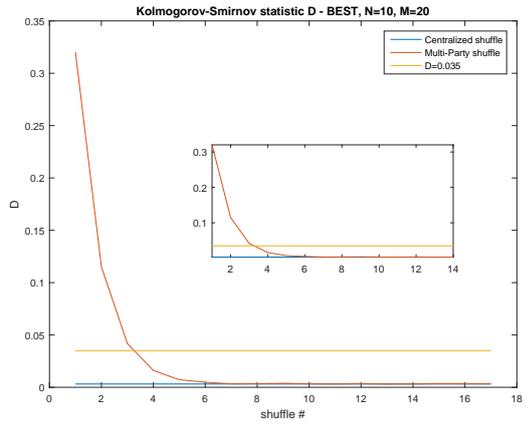}
		\caption{$N=10$, $M=20$}
		\label{fig:ksStatistic_best2}
	\end{subfigure}
	\caption{Kolmogorov-Smirnov test for different selection of $N$ and $M$ of the best case scenario}
	\label{fig:ksStatistic_comparisonBest}
\end{figure}

\begin{figure}[H]
	\centering
	\begin{subfigure}[b]{0.48\textwidth}
		\centering
		\includegraphics[trim=50 190 50 190, clip, width=\textwidth]{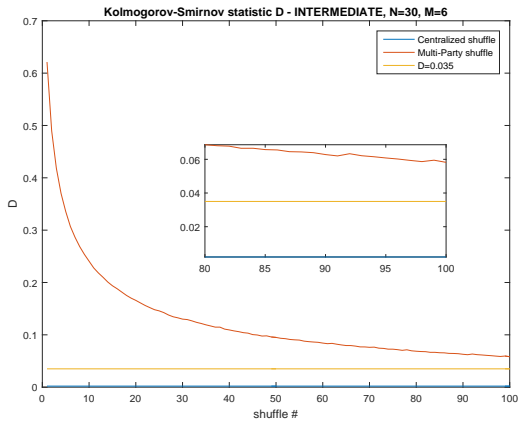}
		\caption{$N=30$, $M=6$}
		\label{fig:ksStatistic_med1}
	\end{subfigure}
	\begin{subfigure}[b]{0.48\textwidth}
		\includegraphics[trim=50 190 50 190, clip, width=\textwidth]{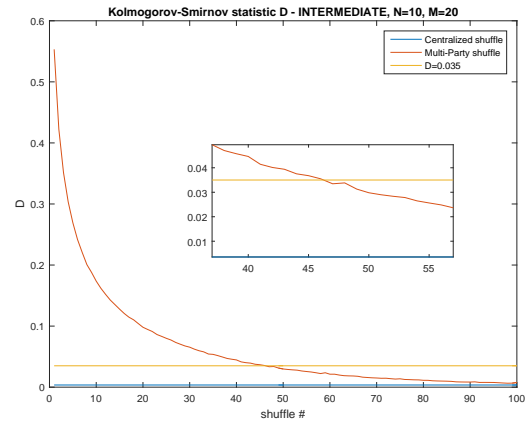}
		\caption{$N=10$, $M=20$}
		\label{fig:ksStatistic_med2}
	\end{subfigure}
	\caption{Kolmogorov-Smirnov test for different selection of N and M of the intermediate case scenario}
	\label{fig:ksStatistic_comparisonMed}
\end{figure}

\begin{figure}[H]
	\centering
	\begin{subfigure}[b]{0.48\textwidth}
		\centering
		\includegraphics[trim=50 190 50 190, clip, width=\textwidth]{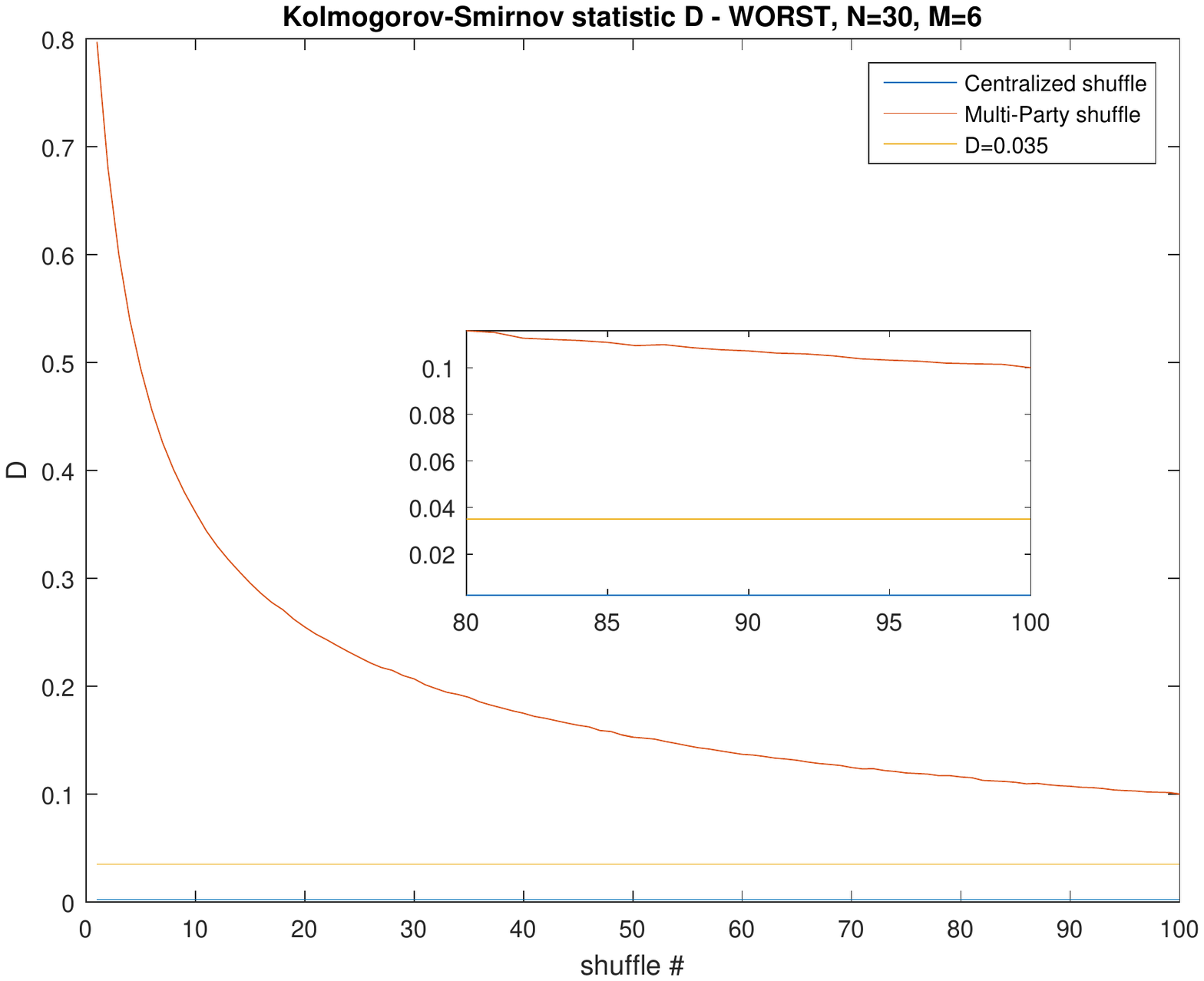}
		\caption{$N=30$, $M=6$}
		\label{fig:ksStatistic_worst1}
	\end{subfigure}
	\begin{subfigure}[b]{0.48\textwidth}
		\includegraphics[trim=50 190 50 190, clip, width=\textwidth]{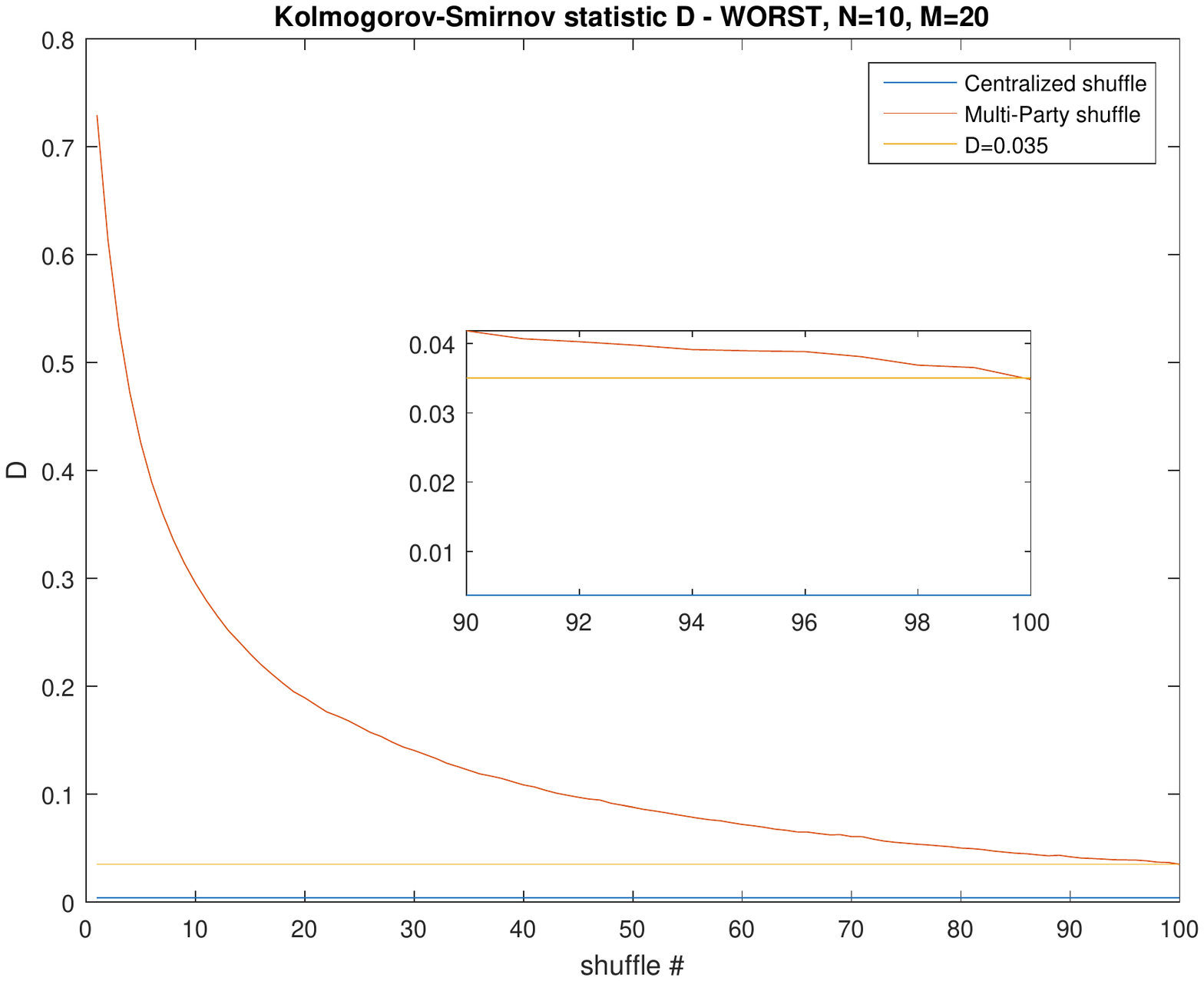}
		\caption{$N=10$, $M=20$}
		\label{fig:ksStatistic_worst2}
	\end{subfigure}
	\caption{Kolmogorov-Smirnov test for different selection of N and M of the worst case scenario}
	\label{fig:ksStatistic_comparisonWorst}
\end{figure}

For the fully connected topology (best case), varying the number of nodes or number of data items does not seem to have an effect on performance. For the line topology (intermediate and worst case), increasing the number of nodes also increases the number of shuffles necessary. However increasing the number of data items does not have a noticeable effect for those cases. These conclusions can be verified in the figures \ref{fig:ksStatistic_comparisonBest},\ref{fig:ksStatistic_comparisonMed}, and \ref{fig:ksStatistic_comparisonWorst} which show the Kolmogorov-Smirnov statistic as function of the shuffles for different selection of total nodes $N$ and total data items $M$. 

\section{Conclusion and Future Work}
\label{sec:oMPSconclusion}
In this paper, we evaluated a basic peer to peer opportunistic mixing strategy in order to generate a baseline of results which can later be used to compare different strategies. We did this by opportunistically shuffling the data among the participants and showed that the number of shuffles is dependent on the properties of the graph that represents the participant interconnections. A fully connected topology requires only $4$ shuffling rounds.
On the other hand, a line topology required significantly more shuffling rounds; $46$ rounds for the intermediate case and $100$ shuffling rounds for the worst case. Using real user GPS traces from the MDC dataset we saw that the number of shuffling rounds did not exceed the worst case when selecting random nodes from the population but at $85$ rounds it was significantly higher than the intermediate case.
Carefully selecting nodes from the population in the MDC dataset to form a more connected topology made a significant difference in the efficiency of the shuffling (only $40$ shuffling rounds) and was significantly better than the intermediate case.

These results can be used to define the stopping criteria for a near uniform shuffle based on the topology of the nodes. A set of $10$ nodes in a fully connected topology (having an average closeness centrality of $1$) would require at least $4$ shuffles. Whereas a set of $10$ nodes in a line topology (having an average closeness centrality of $0.3430$), would require $100$ shuffles. For the MDC dataset the closeness centrality ranged from $0.3430$ to $1$ and from $0.6$ to $1$ for the random user selection and the clique-based user selection respectively. 

Opportunistic peer to peer mixing, as part of a slicing and mixing strategy, can therefore reasonably mix the data so as to protect the identity of the source in the context of the data routing. However, the data content itself should be further obfuscated in order to protect the identity of the source which might be revealed from analyzing the data content. Such techniques require the manipulation of data entries and can reduce the quality of the data but it is often necessary to do so for the protection of the participants.

Based on the positive results of these simulations, we plan to implement this in a real study to verify the results with mobility data that is collected in a different city.

\bibliography{PaperReferences-opportunistic_mps_2017}

\end{document}